\documentclass[prb,aps,onecolumn]{revtex4}
\usepackage{amsmath}
\usepackage{wasysym}
\usepackage[dvips]{graphicx}

\begin{document}
\newcommand{\vl}{v_{_L}}
\newcommand{\vc}{\mathbf}
\newcommand{\be}{\begin{equation}}
\newcommand{\ee}{\end{equation}}
\newcommand{\bk}{{{\bf{k}}}}
\newcommand{\bK}{{{\bf{K}}}}
\newcommand{\cE}{{{\cal E}}}
\newcommand{\bQ}{{{\bf{Q}}}}
\newcommand{\br}{{{\bf{r}}}}
\newcommand{\bg}{{{\bf{g}}}}
\newcommand{\bG}{{{\bf{G}}}}
\newcommand{\hbr}{{\hat{\bf{r}}}}
\newcommand{\bR}{{{\bf{R}}}}
\newcommand{\bq}{{\bf{q}}}
\newcommand{\hx}{{\hat{x}}}
\newcommand{\hy}{{\hat{y}}}
\newcommand{\hz}{{\hat{z}}}
\newcommand{\bea}{\begin{eqnarray}}
\newcommand{\eea}{\end{eqnarray}}
\newcommand{\ra}{\rangle}
\newcommand{\la}{\langle}
\newcommand{\upa}{\uparrow}
\newcommand{\dna}{\downarrow}
\newcommand{\bS}{{\bf S}}
\newcommand{\vS}{\vec{S}}
\newcommand{\dg}{{\dagger}}
\newcommand{\pdg}{{\phantom\dagger}}

\title{N\'eel to staggered dimer order transition in a generalized honeycomb lattice Heisenberg model}
\author{Argha Banerjee}
\affiliation{Tata Institute of Fundamental Research, 1, Homi Bhabha Road, Mumbai, India 400 005}
\author{Kedar Damle}
\affiliation{Tata Institute of Fundamental Research, 1, Homi Bhabha Road, Mumbai, India 400 005}
\author{Arun Paramekanti}
\affiliation{Department of Physics, University of Toronto, Toronto, Ontario M5S 1A7, Canada}
\affiliation{Canadian Institute for Advanced Research, Toronto, Ontario, M5G 1Z8, Canada}

\begin{abstract}
We study a generalized honeycomb lattice spin-1/2 Heisenberg model with nearest-neighbor antiferromagnetic
2-spin exchange, and competing 4-spin interactions which serve to stabilize a staggered dimer state which
breaks lattice rotational symmetry. Using a combination of quantum Monte Carlo numerics, spin
wave theory, and bond operator theory, we show that this model undergoes a strong first-order
transition between a N\'eel state and a staggered dimer state upon increasing the strength of the 4-spin
interactions. We attribute the strong first order
character of this transition to the spinless nature of the core of point-like $Z_3$ vortices obtained in 
the staggered dimer state. Unlike in the case of a columnar dimer state, disordering such vortices in the
staggered dimer state does not naturally lead to magnetic order, suggesting that, in this model, 
the dimer and N\'eel
order parameters should be thought of as independent fields as in conventional
Landau theory.
\end{abstract}
\maketitle
\section{Introduction}
Phase transitions between two phases of matter distinguished by symmetry properties
are usually well-described by Landau theory. In its simplest form, Landau theory
expresses the free energy of the system as an analytic function of the order parameter
whose value captures the symmetry breaking inherent in the ordered state. All analytic terms
consistent with the symmetries of the microscopic Hamiltonian are included in this
free energy function, with coefficients that are undetermined functions of the microscopic
interactions in the system.
The phase of the system in this description is obtained by minimizing this free energy
function over different values of the order parameter. In this description, phase transitions to
the broken symmetry long-range ordered phase are driven by changes in the values of the coefficients of various terms,
which change the
position of the minimum to a non-zero value of the order parameter at the phase transition. Symmetry considerations,
which dictate the form of the terms allowed in the Landau free energy function, then allow one
to decide whether a particular phase transition is generically a first order transition or second-order in nature.
For instance, for systems with a global $Z_2$ symmetry and a scalar order parameter (such as the Ising model), this
Landau theory approach correctly predicts that the transition to the symmetry breaking long-range ordered phase
is generically a second-order transition, with the order parameter growing continuously from zero
at the transition.
On the other hand, when the two phases on either side of the transition break different symmetries, and thereby
possess different order parameters, the prediction of Landau theory in the generic case
is  that the change from one phase to the other proceeds either by via multiple transitions (going through an 
intermediate phase with coexisting orders or with both orders being absent), 
or via a direct first-order transition with one
order parameter abruptly jumping to zero and the other abruptly becoming non-zero at precisely
the same transition point. 

Recent work by Senthil and co-authors~\cite{Senthil_Science_2004,Senthil_etal_2004} suggests that such a 
Landau theory approach is misleading for a class
of quantum phase transitions between Neel ordered antiferromagnets and valence-bond ordered paramagnets, 
most notably
in two-dimensional square-lattice antiferromagnets. In this case, Landau theory would proceed by
writing down the free energy function as an expansion in the Neel order parameter ${\mathbf n}$ and
the valence-bond order parameter $\psi$. Since one of these lives in spin-space, and the other
represents order in real-space, Landau theory considerations would predict a first-order transition
or an intermediate phase in the generic case. Such an intermediate phase
appears likely for instance on the honeycomb lattice,
when an additional next-nearest neighbour exchange
coupling destroys the Neel ordering of the nearest neighbour Heisenberg
antiferromagnet~\cite{Arun}. However, Ref~\onlinecite{Senthil_etal_2004} argues that
the transition can in fact be a generically continuous transition, and is better described in terms
of `deconfined' spinon variables rather than the order parameter fields of Landau theory. This failure
of Landau theory has been ascribed to the presence
of crucial Berry phase terms in the action written in terms of the order parameter fields, and it is
conjectured that when these are correctly taken into account, one arrives naturally at a
description in terms of `deconfined' spinons interacting with a $U(1)$ gauge field. This continuum description leads to a prediction
of a direct second order transition between the two phases if certain
monopole operators allowed in the theory are actually irrelevant at
the fixed point describing the transition, rendering the emergent
gauge field effectively non-compact. This was conjectured to be the case
for the transition from the Neel ordered antiferromagnet to a four-fold symmetry breaking valence bond solid phase
in square lattice antiferromagnets. Various numerical works on a particular
spin model with multiple-spin interactions appears to lend support to these field theoretic ideas.
Furthermore, even in cases where
these monopole operators are not irrelevant, this approach
suggests that the transition would be weakly-first order,  allowing one to use this continuum description to describe the physics at
all but the largest length scales at which the weakly-first order
nature of the transition asserts itself. 

A more intuitive view of `deconfined criticality' was provided in subsequent work by Levin and
Senthil~\cite{Levin_Senthil}
who came up with a simple picture for the spin-1/2 spinon variables and the $U(1)$ gauge
field that make up the basic ingredients for the correct description of a generically continuous
transition between a Neel ordered antiferromagnet and a columnar valence-bond
solid 
(VBS) ordered paramagnet on the square lattice. The basic idea was that a $Z_4$ vortex formed by breaking up the sample into four domains
of solid order meeting at the vortex core necessarily contains a free spin-1/2 variable localized at its
core. The core energy of these vortices is expected to decrease upon approaching
the vicinity of the transition to the antiferromagnetic phase. Under the assumption that the $Z_4$
anisotropy is irrelevant, the phase of the order parameter winds continuously around the core
(as in $U(1)$ vortices). It is then
natural to think of this transition to the antiferromagnet in terms of the proliferation of these $U(1)$ vortices,
and write down a theory for the transition in these vortex variables.
Since solid order is destroyed by the proliferation of these vortices, the destruction
of the solid order is accompanied by establishment of spin ordering corresponding to
the condensation of these spin-1/2 degrees of freedom in the cores of the proliferating
vortices. Reasoning in this manner, Levin and Senthil were able to deduce the form
of the continuum theory for the resulting phase transition from
symmetry arguments and these simple intuitive considerations.


Motivated by this simple argument, it seems reasonable to make the following supposition:
If the lowest core-energy vortices in the valence-bond order have a free spin in their cores,
then the transition to the adjacent antiferromagnetic phase may be expected
to admit a natural description
in spinon variables, of the type developed in Ref.~\onlinecite{Senthil_etal_2004}.\begin{figure}[h]
{\includegraphics[width=\columnwidth]{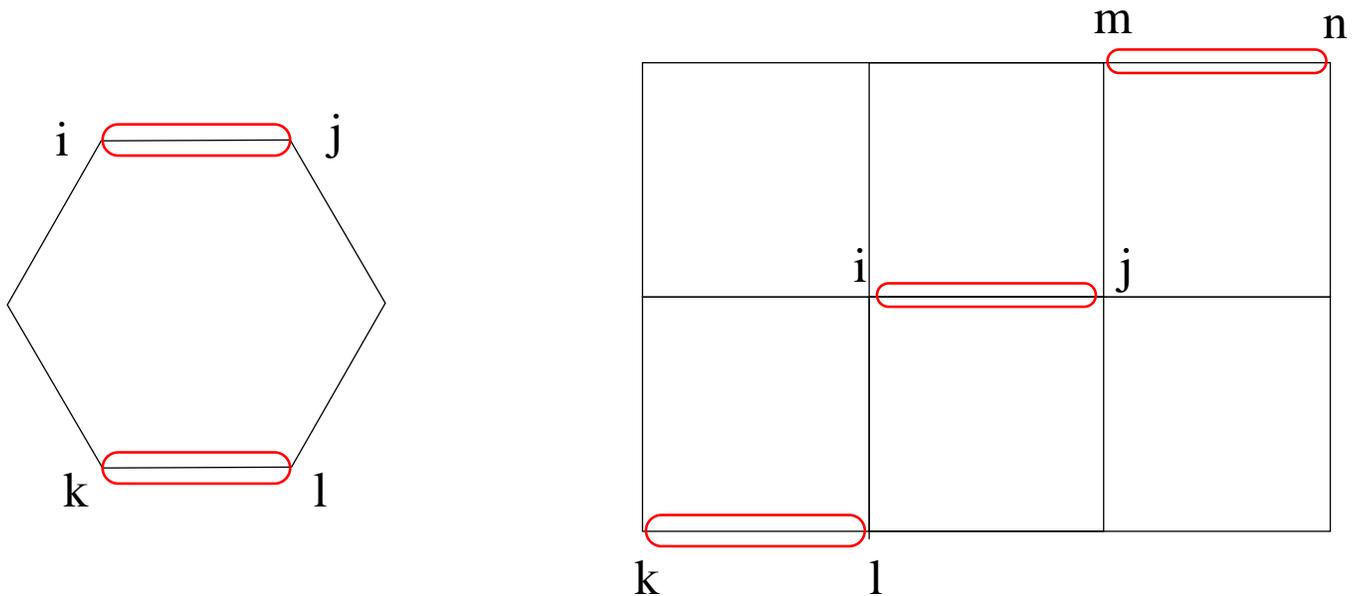}}
\caption{A four-spin operator in the  $JQ$ model on the honeycomb lattice (left panel)
and a six-spin operator (right panel)in the staggered $JQ_3$ model  on the square lattice.} 
\label{plaqs_stg}
\end{figure}
This would imply either a generically continuous phase transition if the $Z_n$ spatial anisotropy 
(with $n=4$ for the square lattice) is
irrelevant at the critical fixed point,
or a weakly-first order transition if this anisotropy turn out to be relevant.
On the other hand, if the lowest core-energy vortices have no free spin
in their core, there is no natural way to obtain magnetic order by proliferating vortices in the
dimer order; one may then expect standard Landau theory to be valid, and the transition
to a nearby antiferromagnetic state would then proceed via an intermediate phase or
be strongly first order.\begin{figure}[h]
{\includegraphics[width=\columnwidth]{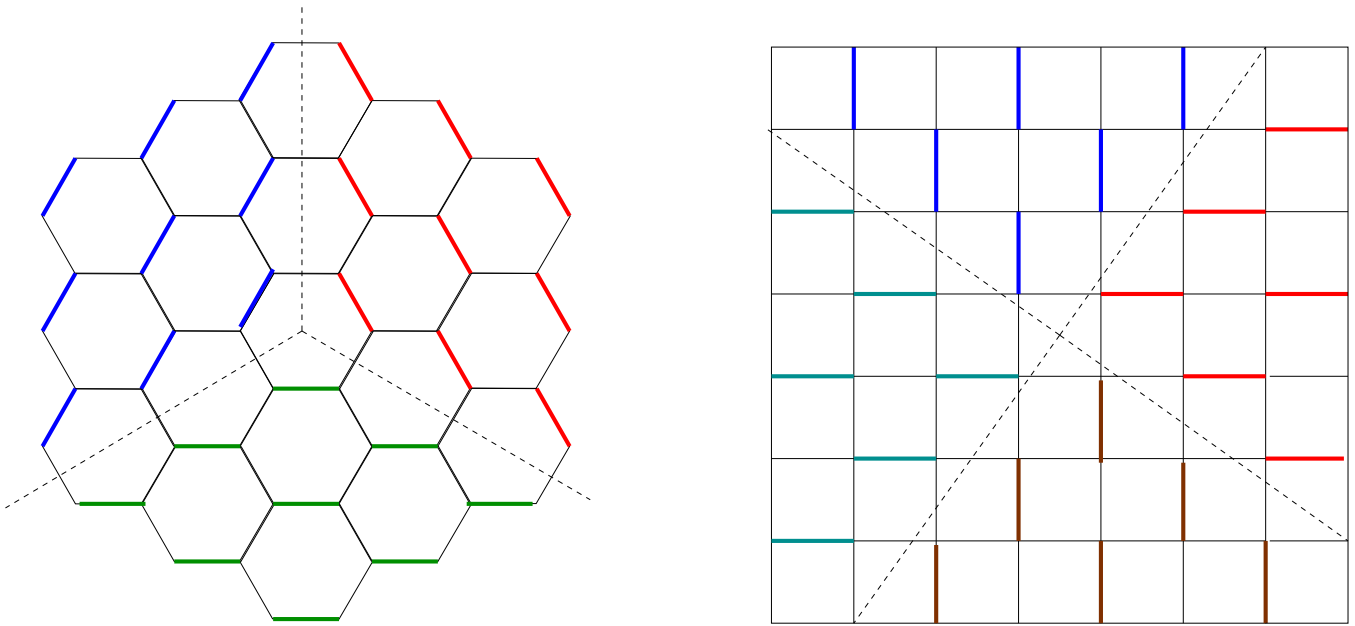}}
\caption{Cartoon of vortices in staggered valence bond nematic on the honeycomb lattice (left panel)
and the staggered valence bond solid (right panel) on the square lattice. Note that one is not
forced to have a free spin in the core of these vortices.} 
\label{vortex_stg}
\end{figure}

Some evidence for this point of view has already been provided by recent work on
models that exhibit a strongly first order transition between a `staggered' valence bond
solid order and Neel order on the square lattice~\cite{Sen_Sandvik_2010,Banerjee_etal_unpublished} in a model with nearest neighbour exchange and competing
6 spin interactions on the square lattice:
\begin{equation}
\label{JQ3_staggered}
H = -J \sum_{\langle ij \rangle} P_{ij} - Q_{3}\sum_{\langle ij, kl, mn \rangle}P_{ij}P_{kl}P_{mn},
\end{equation}
where $P_{ij}$ denotes a bipartite singlet projector, 
$P_{ij} = \frac{1}{4} - {\mathbf S}_{i} \cdot {\mathbf S}_{j}$.
The plaquette interactions $Q_3$ in the formula above are represented pictorially in
Fig.~\ref{plaqs_stg} (right panel).

\begin{figure}[h]
{\includegraphics[width=\columnwidth]{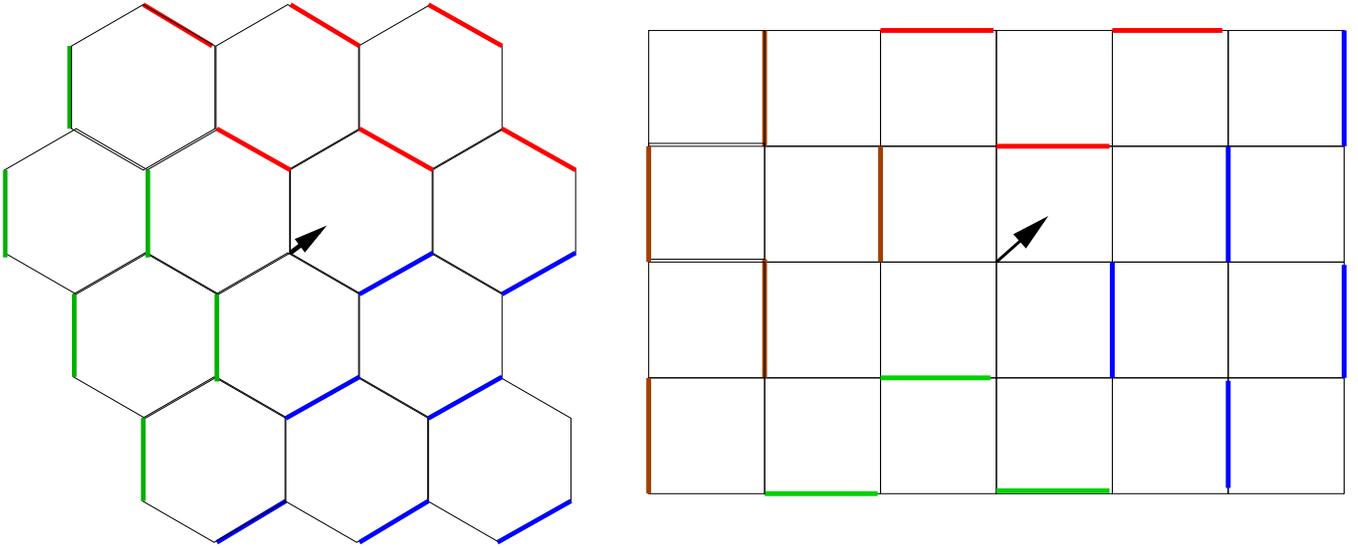}}
\caption{Vortices with free spins in their cores are also possible
in staggered valence bond nematic on the honeycomb lattice (left panel)
and the staggered valence bond solid (right panel) on the square lattice, but are expected
to cost higher core energy than ones with spinless cores.} 
\label{vortex_stg_spinful}
\end{figure}

As is clear from Fig~\ref{vortex_stg} (right panel), the
simplest caricatures of vortices in the staggered valence bond solid
order on the square lattice are indeed without any free spins at their core.
In the staggered case, it is also possible to construct vortices with spins
at their cores, but these are expected to cost more energy due
to additional singlet formation cost associated with leaving
one spin free in the core, as is clear from Fig~\ref{vortex_stg_spinful} (right panel). This suggests that the 
transition would be strongly first order, which is consistent
with recent numerical works~\cite{Sen_Sandvik_2010,Banerjee_etal_unpublished}. This should be contrasted 
with studies of very similar ``$JQ$'' models on the square lattice, with the plaquette interactions chosen to 
favour columnar order\cite{Sandvik_PRL_2007}. Extensive numerical work on these models has led to the conclusion
that the transition from N\'eel to columnar valence bond solid order is indeed a continuous 
transition,~\cite{Lou_Sandvik_Kawashima,Melko_Kaul,Sandvik_PRL_2010} although the critical point exhibits 
logarithmic violations of standard finite size scaling~\cite{Sandvik_PRL_2010,Banerjee_Damle_Alet}. These violations
of scaling have also been interpreted  as possible evidence of first order behaviour~\cite{Jiang_et_al}. $SU(3)$ and $SU(4)$
versions of the same transition have also been studied~\cite{Lou_Sandvik_Kawashima,Kaul,Banerjee_Damle_Alet_SU3} and found to be continuous in nature.   
 
Here we focus on a related system on the honeycomb lattice: On the honeycomb lattice,
the staggered ordering of near-neighbour valence bonds does not break lattice translation symmetry,
but does break symmetry of three-fold lattice rotations. As is clear from
Fig~\ref{vortex_stg} (left panel), the simplest construction of elementary
$Z_3$ vortices in this valence-bond nematic
order naturally yields vortices with spinless cores. Again, vortices with spinful cores are also possible, but 
are expected to cost more core-energy due to two reasons. First, forming a spinful vortex core 
shown in Fig~\ref{vortex_stg_spinful} (left panel) involves
the breaking of additional singlet bonds, and would cost an additional core energy of the order of the spin gap.
At the same time, the spinless vortex core, shown in Fig.~\ref{vortex_stg} (left panel), allows for 
dimer resonances on the core plaquette which is expected to lead to core energy lowering.

The transition
between such a valence-bond nematic and a nearby antiferromagnetic phase on the honeycomb lattice is
thus expected to be strongly first order based on our supposition above. Here we focus
on a spin Hamiltonian with nearest neighbour exchange couplings that compete with four-spin interaction
terms that stabilize a valence-bond nematic state. We study this system numerically by a projector Monte Carlo
algorithm in the valence-bond basis, as well as analytically by spinwave and bond-operator expansions.
Our numerics finds a strongly first order transition between a valence-bond nematic and a antiferromagnet,
consistent with the intuitive picture outlined above. In addition, we use our approximate
analytical calculations to provide a reasonable semi-quantitative account of our
numerical results. We note that an exactly soluble model with such a valence bond nematic ground
state has also been proposed recently.\cite{brijesh2010}

The remainder of the paper is organized as follows: In section~\ref{model}, we define
the model studied, and describe the approximate analytical methods that we
use for their study. In section~\ref{numerics}, we describe our
numerical approach and summarize the results obtained. Finally,
we conclude with a brief discussion in section~\ref{discussion}.

\section{Model and analytical treatment}
\label{model}

The honeycomb lattice staggered $JQ$ model Hamiltonian is defined as folows,
\begin{equation}
\label{JQ2_staggered}
H = -J \sum_{\langle ij \rangle} P_{ij} - Q\sum_{\langle ij, kl \rangle}P_{ij}P_{kl},
\end{equation}
where the four spin plaquette operators are products of two bipartite singlet projectors
acting on  two parallel bonds of the hexagons in the honeycomb lattice 
(see Fig.~\ref{plaqs_stg}(left panel)). 
This Hamiltonian can be written explicitly as a sum of 
two-spin and four-spin interactions as follows,   
\bea
H_{\rm JQ} &=& H_2 + H_4 \\
H_2 &=& \tilde{J} \sum_{\langle i,j \rangle} \bS_i\cdot\bS_j \\
H_4 &=& - Q \sum_{\hexagon}
\left[ (\bS^{\hexagon}_1 \cdot\bS^{\hexagon}_2)
(\bS^{\hexagon}_4 \cdot\bS^{\hexagon}_5)
+ (\bS^{\hexagon}_2 \cdot\bS^{\hexagon}_3)
(\bS^{\hexagon}_5 \cdot\bS^{\hexagon}_6) +
(\bS^{\hexagon}_3 \cdot\bS^{\hexagon}_4)
(\bS^{\hexagon}_6 \cdot\bS^{\hexagon}_1) \right]
\eea
where $\sum_{\hexagon}$ denotes  a sum over all hexagons of the lattice, $\bS^{\hexagon}_i$
denotes the $i$-th spin on a hexagon, with
$i=1$-$6$ labelling the 6 sites clockwise around the hexagon, and $\tilde{J}$ = $J +Q/2$.

At a heuristic level, we expect that the $Q$ term drives a transition to a valence bond
nematic state with staggered ordering of near-neighbour singlet bonds, while the $\tilde{J}$
term favours a Neel state with spins on the $A$-sublattice ($B$-sublattice) of the honeycomb lattice
aligned (anti-aligned) along a spontaneously chosen axis in spin space. A simple approximate description of the N\'eel
phase is obtained by doing a standard spin wave analysis. On the other hand, to describe the valence-bond
nematic, we use a description in terms of bond-operators that becomes very accurate
in the limit of strong nematic order. By comparing the predictions
for the ground state energy from these two approximate calculations, we also estimate the value
of $Q$ at which we expect a transition between the two phases.

\subsection{Spin wave theory of the Neel state}
In the N\'eel ordered phase of the JQ model defined above,
let us assume we can decouple the four spin interactions as
\be
(\bS^{\hexagon}_1 \cdot\bS^{\hexagon}_2) (\bS^{\hexagon}_4 \cdot\bS^{\hexagon}_5)
= -\alpha (\bS^{\hexagon}_1 \cdot\bS^{\hexagon}_2 + \bS^{\hexagon}_4 \cdot\bS^{\hexagon}_5) - \alpha^2
\ee
where $\alpha \equiv - \la \bS_i \cdot\bS_j\ra$ is the nearest neighbor spin correlation which is 
assumed to be the same on all neighboring bonds and
which needs to be determined self-consistently. Such a decomposition can alternatively be  obtained within
a path integral formalism by using
a Hubbard Stratonovich (HS) decoupling of the quartic spin operators followed by a static mean field
theory of the HS field.
In this case, the Hamiltonian reduces to the form
\be
H_{\rm mf} = J_{\rm eff} \sum_{\la i,j\ra} \bS_i \cdot \bS_j + 3 Q  \alpha^2 N_H
\label{hmf}
\ee
where we have defined $J_{\rm eff}=\tilde{J}+2 Q \alpha$, and 
$N_H$ is the number of unit cells on the lattice 
(the number of spins is $N_S = 2 N_H$).

We treat this effective Hamiltonian using the standard
spin wave theory by defining
Holstein Primakoff bosons via
$(S_A^z,S_A^+,S_A^-) \equiv (\frac{1}{2}-a^\dg a^\pdg,a^\pdg,a^\dg)$ and $(S_B^z,S_B^+,S_B^-) 
\equiv (b^\dg b^\pdg - \frac{1}{2},b^\dg,b^\pdg)$ for sites on the A-sublattice and
B-sublattice respectively. 
Retaining terms to quadratic order in the bosons leads to the Hamiltonian
\be
H_{\rm mf}  =
\frac{J_{\rm eff}}{2} \sum_\bk \begin{pmatrix} a^\dg_\bk & b^\pdg_{-\bk}  \end{pmatrix} \\
\begin{pmatrix} 3 & \Gamma_\bk \\ \Gamma^*_\bk & 3 \end{pmatrix} \\
\begin{pmatrix} a^\pdg_\bk \\ b^\dg_{-\bk}  \end{pmatrix} \\ - \frac{3}{4} J_{\rm eff} N_H + 3 Q \alpha^2 N_H
 - \frac{J_{\rm eff}}{2} \sum_\bk 3
\ee
where 
$\Gamma_\bk \equiv (1 + {\rm  e}^{-i k_b} + 
{\rm  e}^{-i k_a- k_b})$ with $k_a=\bk\cdot\hat{a}$ and
$k_b=\bk\cdot\hat{b}$. Setting $\Gamma_\bk = |\Gamma_\bk| {\rm e}^{i \gamma_\bk}$, we can diagonalize this
Hamiltonian via a Bogoliubov transformation
\be
\begin{pmatrix} a^\pdg_\bk \\ b^\dg_{-\bk} \end{pmatrix}
= \begin{pmatrix} \cosh\theta_\bk {\rm e}^{i\gamma_\bk} & -\sinh \theta_\bk \\
-\sinh\theta_\bk & \cosh\theta_\bk {\rm e}^{-i\gamma_\bk} \end{pmatrix}
\begin{pmatrix} c^\pdg_\bk \\ d^\dg_{-\bk} \end{pmatrix}
\ee
where 
$\sinh 2\theta_\bk = |\Gamma_\bk|/\Omega_\bk$,
$\cosh 2\theta_\bk =3/\Omega_\bk$, and $\Omega_\bk=\sqrt{9 - |\Gamma_\bk|^2}$.
This yields the diagonalized Hamiltonian
\bea
H_{\rm mf}  &=& - \frac{3}{4} J_{\rm eff} N_H + 3 Q \alpha^2 N_H + \frac{J_{\rm eff}}{2}  \sum_\bk (\Omega_\bk - 3)
+
\frac{J_{\rm eff}}{2} \sum_\bk \Omega_\bk (c^\dg_\bk c^\pdg_\bk + d^\dg_\bk d^\pdg_\bk)
\eea
with a ground state energy
\be
E_{\rm SW} = - \frac{3}{4} J_{\rm eff} N_H + 3 Q \alpha^2 N_H + \frac{J_{\rm eff}}{2}  \sum_\bk (\Omega_\bk - 3)
\label{esw}
\ee
Applying the Feynman-Hellman theorem to the mean field Hamiltonian, $H_{\rm mf}$ in Eq.\ref{hmf}, yields
\be 
\alpha = - \frac{1}{3 N_H} \frac{\partial E_{\rm SW}}{\partial J_{\rm eff}} = 
\frac{1}{4} - \frac{1}{6 N_H} \sum_\bk (\Omega_\bk - 3)
\label{alpha}
\ee
Eqns.~(\ref{esw}) and (\ref{alpha}), together, determine the ground state energy of this model in the N\'eel ordered phase. A feature of our mean-field spin-wave analysis is that 
$\alpha$ is independent of $Q/\tilde{J}$.  Numerically, we find $\alpha \approx 0.3549$ within this spin wave approach which agrees with 
 earlier spin wave results \cite{honeycombMC}.
This value is quite close to the value, $\alpha \approx 0.3627$, deduced from a recent optimized valence bond trial
wave function study \cite{jafari2010}
of the nearest neighbor Heisenberg model, and 
series expansion results \cite{honeycombseries}
on the Heisenberg model which give $\alpha \approx 0.3659$. This spin wave
result for $\alpha$ must be
used as input to compute the ground state energy per spin of the $J$-$Q$ model which is given by
$e_{\rm SW} = -\frac{3}{2} (\tilde{J} + Q \alpha) \alpha$; this is plotted in Fig.\ref{fig:energy}. 
Within this approach, the ordered moment is also independent
of $Q/\tilde{J}$, and is given by
\be
m = \frac{1}{2} - \frac{1}{2 N_H} \sum_\bk (\frac{3}{\Omega_\bk}-1)
\ee
and takes on a value $m \approx 0.2420$, whereas series expansion studies of the
nearest neighbor Heisenberg model \cite{honeycombseries}
yield $m \approx 0.266$.
In the next section, we will check to what extent the numerical results bear
out our prediction of a nearly $Q$-independent value of $m$ in the N\'eel phase.

\subsection{Bond operator theory of the staggered dimer state}
In the other limit, when the $Q$ term dominates over $\tilde{J}$ Hamiltonian at large $Q$ has spontaneous staggered dimer order in the
ground state, we can 
use a bond operator formalism to compute the ground state energy and correlations in this phase. 
We label
the spins by their unit cell position $\br = m\hat{a} + n \hat{b}$ and a sublattice index $p = 1,2$.
In the staggered dimer state, let us assume that the spins at sites $(\br,1)$ and $(\br,2)$ in every unit
cell form a singlet. This dimer pattern then spontaneously breaks the lattice rotational symmetry but
leaves the translational symmetry intact. 
We define the singlet and triplet states on this pair of sites in terms of bond operators via
\bea
|s\ra &=& \frac{1}{\sqrt{2}} (\upa_1 \dna_2 - \dna_1\upa_2) \equiv s^\dagger |0\ra \\
|z\ra &=& \frac{1}{\sqrt{2}} (\upa_1 \dna_2 + \dna_1\upa_2)  \equiv t_z^\dagger |0\ra \\
|x\ra &=& \frac{-1}{\sqrt{2}} (\upa_1 \upa_2 - \dna_1\dna_2)  \equiv t_x^\dagger |0\ra \\
|y\ra &=& \frac{i}{\sqrt{2}} (\upa_1 \upa_2 + \dna_1\dna_2)  \equiv t_y^\dagger |0\ra
\eea
where we must satisfy the constraint $s^\dg_\br s_\br^\pdg + t_{\br\alpha}^\dg t_{\br\alpha}^\pdg =1$ at each site.
The spin operators can be rewritten in terms of the bond operators as
\be
\bS^\alpha_{\br,p} = \frac{1}{2} (-1)^p \big[ s_{\br}^\dg t^\pdg_{\br\alpha} + t_{\br\alpha}^\dg s^\pdg_\br\big]
- \frac{i}{2} \epsilon_{\alpha\beta\gamma} t_{\br\beta}^\dg t_{\br\gamma}^\pdg
\ee
where $\alpha=x,y,z$.
Assuming the staggered dimer state corresponds to a uniform singlet condensate, we can replace 
$s^\dg_\br \to \bar{s}$ and $s^\pdg_\br \to \bar{s}$. Furthermore, if the singlet condensate is robust, so that
the dimer order is strong as observed in the QMC numerics, 
we expect the triplet density to be small and a Bogoliubov-type theory, where the
triplet density plays the role of a small parameter, to be a reasonable starting point. Expanding out the
Hamiltonian in terms of these bond operators, and retaining leading quadratic terms in the triplet operators,
we find
\bea
H_{\rm BO}\!\! &\!\!=\!\!&\!\! 
(\frac{\tilde{J}}{4}\!-\!\lambda) \sum_\br t^\dg_{\br\alpha} t^\pdg_{\br\alpha} + \frac{3 Q}{16} \bar{s}^2 \sum_\br (t^\dg_{\br\alpha} t^\pdg_{\br\alpha} +t^\dg_{\br+\hat{a},\alpha} 
t^\pdg_{\br+\hat{a},\alpha})
\nonumber \\
 &-& \tilde{J} \frac{\bar{s}^2}{4} \sum_{\br} (t^\dg_{\br\alpha} \!+\! t^\pdg_{\br\alpha}) (
t^\dg_{\br-\hat{b},\alpha} \!+\! t^\pdg_{\br-\hat{b},\alpha} \!+\!
t^\dg_{\br-\hat{a}-\hat{b},\alpha} \!+\! t^\pdg_{\br-\hat{a}-\hat{b},\alpha}) \nonumber\\
&-& \frac{3}{4} \tilde{J} \bar{s}^2 N_H - \frac{9}{16} Q \bar{s}^4 N_H - \lambda N_H (\bar{s}^2-1)
\eea
where $\lambda$ is a chemical potential which serves to satisfy the constraint equation on average by
fixing $\la t_{\br\alpha}^\dg t_{\br\alpha}^\pdg\ra  = 1- \bar{s}^2$.
Going to momentum space, the Hamiltonian takes the form
\be
H_{\rm BO} = E^{(0)} + \sum_{\bk}^\prime
\begin{pmatrix} 
t^\dg_{\bk\alpha} & t^\pdg_{-\bk\alpha}
\end{pmatrix} 
\begin{pmatrix} 
A + B_\bk & B_\bk \\
B_\bk & A + B_\bk
\end{pmatrix} 
\begin{pmatrix} 
t^\pdg_{\bk\alpha} \\
t^\dg_{-\bk\alpha}
\end{pmatrix} - \frac{3}{2} A N_H
\ee
where
\bea
E^{(0)} &=& -(\frac{3 \tilde{J}}{4} \bar{s}^2 + \frac{9 Q}{16} \bar{s}^4) N_H 
-\lambda N_H (\bar{s}^2 - 1) \\
A &=& (\frac{\tilde{J}}{4} + \frac{3 Q}{8} \bar{s}^2 - \lambda) \\
B_\bk &=& -\frac{\tilde{J}}{2} \bar{s}^2 (\cos k_b + \cos (k_a +k_b))
\eea
and the prime on the momentum sum indicates that we only sum over
half the Brillouin zone (keeping states with $k_a>0$ for example).
Diagonalizing this via a Bogoliubov transform
\be
\begin{pmatrix}
t^\pdg_{\bk\alpha} \\ t^\dg_{-\bk\alpha} 
\end{pmatrix}
=
\begin{pmatrix}
\cosh\theta_\bk & -\sinh\theta_\bk \\
-\sinh\theta_\bk & \cosh\theta_\bk
\end{pmatrix}
\begin{pmatrix}
c^\pdg_{\bk\alpha} \\ d^\dg_{-\bk\alpha} 
\end{pmatrix}
\ee
we find that the diagonal Hamiltonian takes the form
\be
H = E^{(0)} + \sum_{\bk}^\prime \omega_\bk 
(c^\dg_{\bk\alpha} c^\pdg_{\bk\alpha} + d^\dg_{\bk\alpha} d^\pdg_{\bk\alpha} + 3) - \frac{3}{2} N_H (\frac{\tilde{J}}{4} + \frac{3 Q}{8} \bar{s}^2)
\ee
where $\omega_\bk\equiv \sqrt{A (A + 2 B_\bk)}$.
This leads to a ground state energy
\be
E_{\rm BO} = E^{(0)} + 3 \sum_{\bk}^\prime \omega_\bk - \frac{3}{2} N_H (\frac{\tilde{J}}{4} + \frac{3 Q}{8} \bar{s}^2).
\ee
This energy must be minimized with respect to $\bar{s}$ once $\lambda$ is determined by the number equation
\be
\frac{3}{2 N_H} \sum_\bk (\frac{A + B_\bk}{\omega_\bk} - 1) = 1-\bar{s}^2
\ee
which follows from demanding $\partial E_{\rm BO}/\partial \lambda=0$ in order to satisfy the constraint relation
$\la t_{\br\alpha}^\dg t_{\br\alpha}^\pdg\ra  = 1- \bar{s}^2$.

\begin{figure}[t]
\includegraphics[width=4.5in]{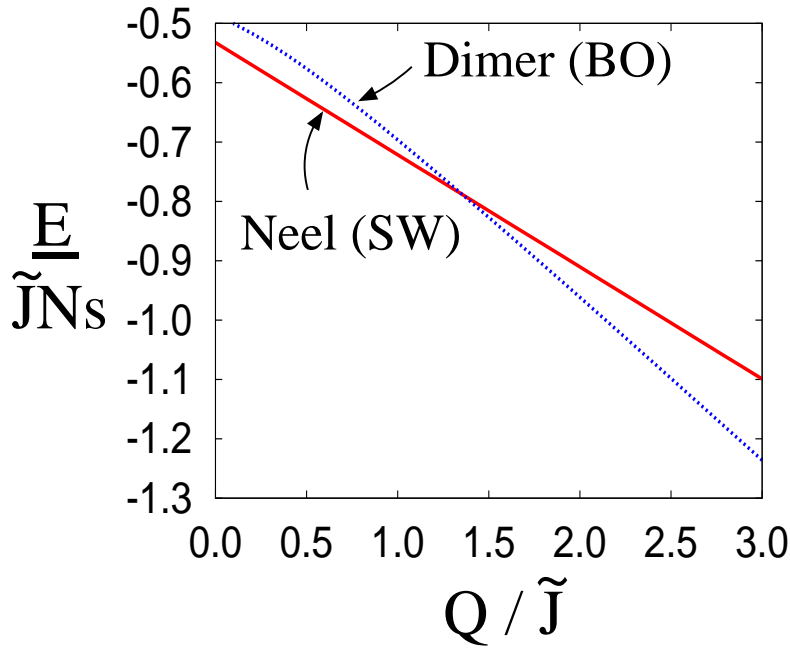}
\caption{(Color online)
Energy per spin (in units of $\tilde{J}$) of the N\'eel state obtained in spin wave theory (SW, red-solid line) compared with
the staggered dimer ordered state obtained using bond operator theory (BO, blue-dotted  line). There is a strong first
order transition at $Q/\tilde{J} \approx 1.35$ $(Q/J \approx 4.2)$.}
\label{fig:energy}
\end{figure}

Solving these equations numerically, we self-consistently determine $\bar{s}$ and $\lambda$. Using these,
we compute the ground state energy and excitation spectrum of the valence bond nematic state. As seen
from Fig.~\ref{fig:energy}, we find that
the energy of this state lies below the antiferromagnetically ordered state for large $Q$, as we expect, but
there is an energy level crossing at $Q/\tilde{J} \approx 1.35$ $(Q/J \approx 4.2)$, so that the N\'eel state
has lower energy at smaller $Q$.
(Within the
bond operator approach, we find that the triplons condense around $Q/\tilde{J} \approx 0.2$ which is well below
the first order transition point.)
Close to the transition, in the spin gapped phase, we find the spin-spin correlation on the dimerized bond to
be $\approx -0.73$ which corresponds to nearly complete dimerization, so that this is a very strong first
order transition. We next turn to a numerically exact quantum Monte Carlo study of this model.

\section{Numerical study}
\label{numerics}

\begin{figure}[h]
{\includegraphics[width=\columnwidth]{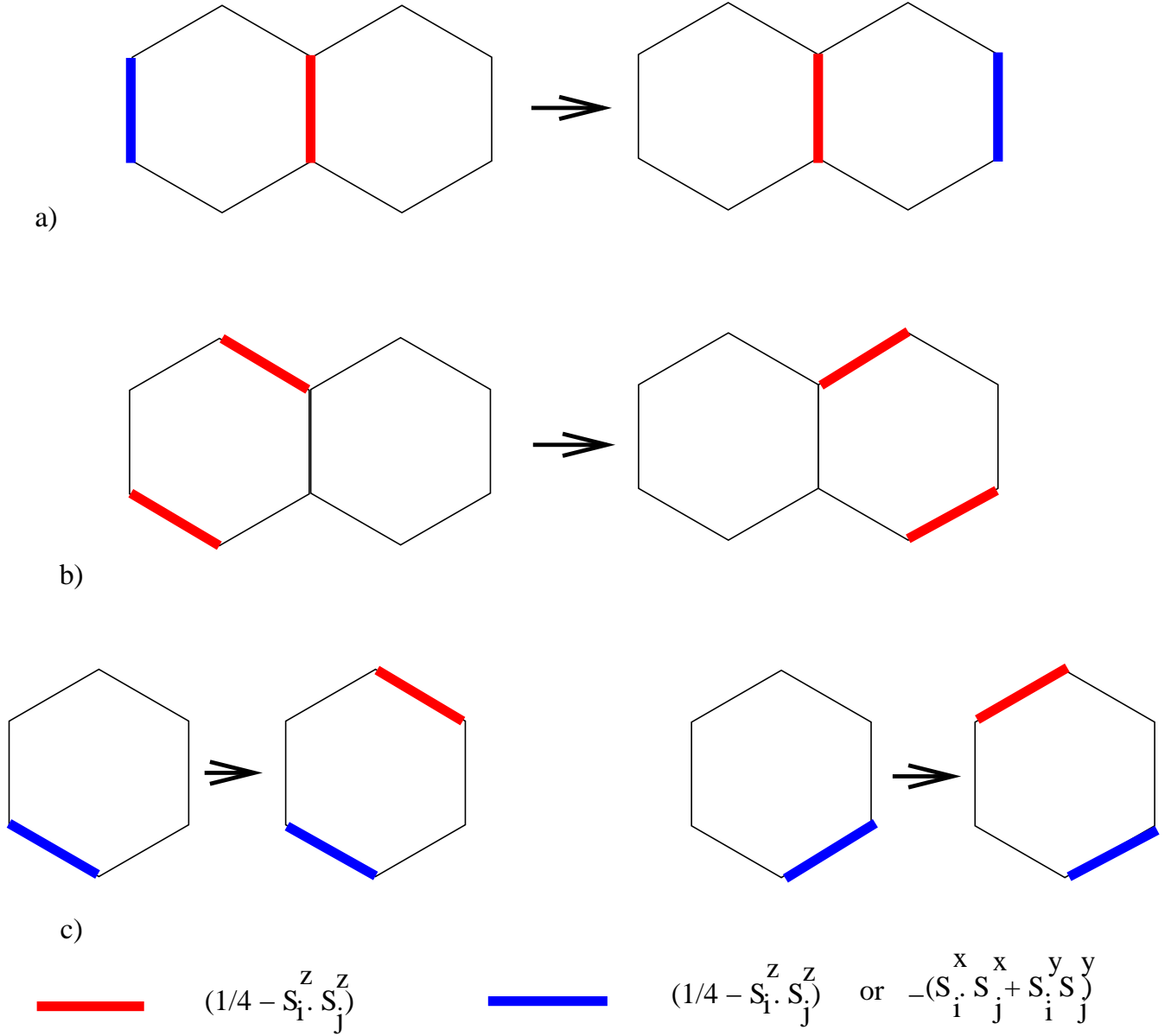}}
\caption{Monte Carlo moves to update plaquette operators: a) reflecting a plaquette operator 
about a bond with a diagonal operator acing on it, b) rotating a diagonal plaquette operator 
about a site, and c) transforming a diagonal or off-diagonal bond operator into a plaquette operator and vice versa.}
\label{updates}
\end{figure}

We use the  valence bond projector Monte Carlo technique~\cite{Sandvik_PRL_2005}  to study   
various ground state properties of the staggered version $JQ$ model on honeycomb lattice as a function of coupling
$Q/J$. Taking advantage of  the improved loop-update technique developed in Ref.~\onlinecite{Sandvik_Evertz},
we scan the phase diagram of this model for systems with upto $2\times32\times32$ sites.  A Monte Carlo projection length of  $6 L^3$ is used to ensure the convergence of 
observables to the ground  state expectation values (we have checked the convergence by comparing with the results of exact diagonalization studies at small sizes). Also, to counter  ergodicity problems in the Monte Carlo simulations 
in the valence bond nematic phase and in the vicinity of the first order transition, we employ many different Monte Carlo moves to update the bond and plaquette operators in a 
Monte Carlo configuration. These updates include attempts to reflect a plaquette operator about its diagonal bond~(Fig.\ref{updates}a),
to rotate the diagonal plaquette operators about a site~(Fig.\ref{updates}b), to transform a plaquette operator into a bond operator by 
deleting a bond and vice versa~(Fig.\ref{updates}c), and so on.
    
We find it convenient to define the valence bond nematic order parameter, 
$\psi(\mathbf{r}) = \mathbf  {S}(\mathbf r)\cdot \mathbf {S}(\mathbf r  + \mathbf e_1 ) + \exp{(i 2 \pi/3)} \mathbf  {S}(\mathbf r)\cdot \mathbf {S}(\mathbf r  + \mathbf e_2 )+  \exp{(i 4 \pi/3)} \mathbf  {S}(\mathbf r)\cdot \mathbf {S}(\mathbf r  + \mathbf e_3 )$, 
where $\mathbf{e_{1,2,3}}$ denote the three nearest neighbor bond vectors of the honeycomb lattice. 
By monitoring the correlations of $\psi$ simultaneously with
those of the spins, we are able to distinguish easily between the
Neel antiferromagnet, and the rotation-symmetry
breaking valence bond nematic. To do this we measure  
\begin{equation}
C(\mathbf{l})= \frac{1}{2N_s}\sum_{{\mathbf r} \in A}\langle \mathbf  {n}(\mathbf r)\cdot \mathbf {n}(\mathbf r  + \mathbf l ) \rangle, 
\end{equation}
where $\mathbf{n}(\mathbf r)= \bf{S}(\mathbf r)- \bf{S}({\mathbf r}+ {\mathbf e_1})$, and the sum over $\bf{r}$ runs over
all $A$-sublattice sites of the lattice. We also define
\begin{equation}
D(\mathbf{l})= \frac{1}{N_s}\sum_{{\mathbf r}}\langle \psi^{*}(\mathbf{r}) \psi(\mathbf{r+l})\rangle,
\end{equation}
where ${\mathbf r}=(x,y)$, and ${\mathbf l}=(L/2,L/2)$ with $L$ being  the linear dimension of the system and $N_s=2L^2$ being the number of sites in the system. 
In the presence of long range antiferromagnetic order, the asymptotic large size limit
of $C(\mathbf{l})$ tends to a finite value, which equals the square of the staggered magnetisation 
\be
{\mathbf M}^2 = \frac{1}{N_s^2}
\sum_{{\bf r}\in A, {\bf r'}\in A} \langle \mathbf  {n}(\mathbf r)\cdot \mathbf {n}(\mathbf r') \rangle.
\ee
In the absence of long range
antiferromagnetic order, we expect this quantity to vanish in the large size limit.
Similarly,  $D(\mathbf{l})$  serves as a order parameter for long range 
valence bond nematic order.

\subsection{Results for honeycomb lattice model}

\begin{figure}[h]
{\includegraphics[width=\columnwidth]{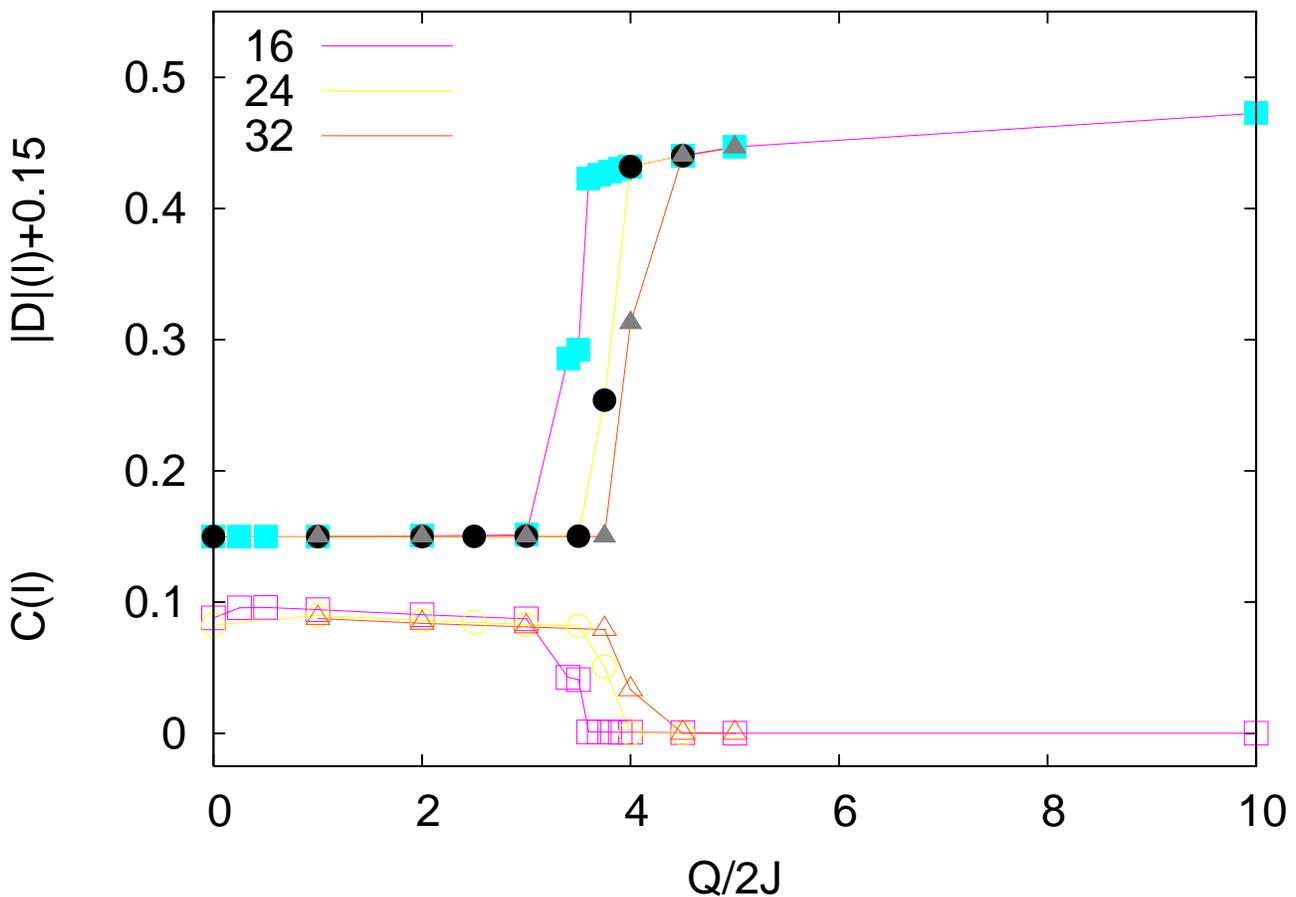}}
\caption{Simultaneous first order jump in N\'eel order parameters $C(\mathbf{l})$ and valence bond nematic order parameter $D(\mathbf{l})$, as $Q/J$ is  varied for the $JQ$ model on the honeycomb lattice.} 
\label{orderparameterVsQ_typ0_stg}
\end{figure}

\begin{figure}[h]
{\includegraphics[width=\columnwidth]{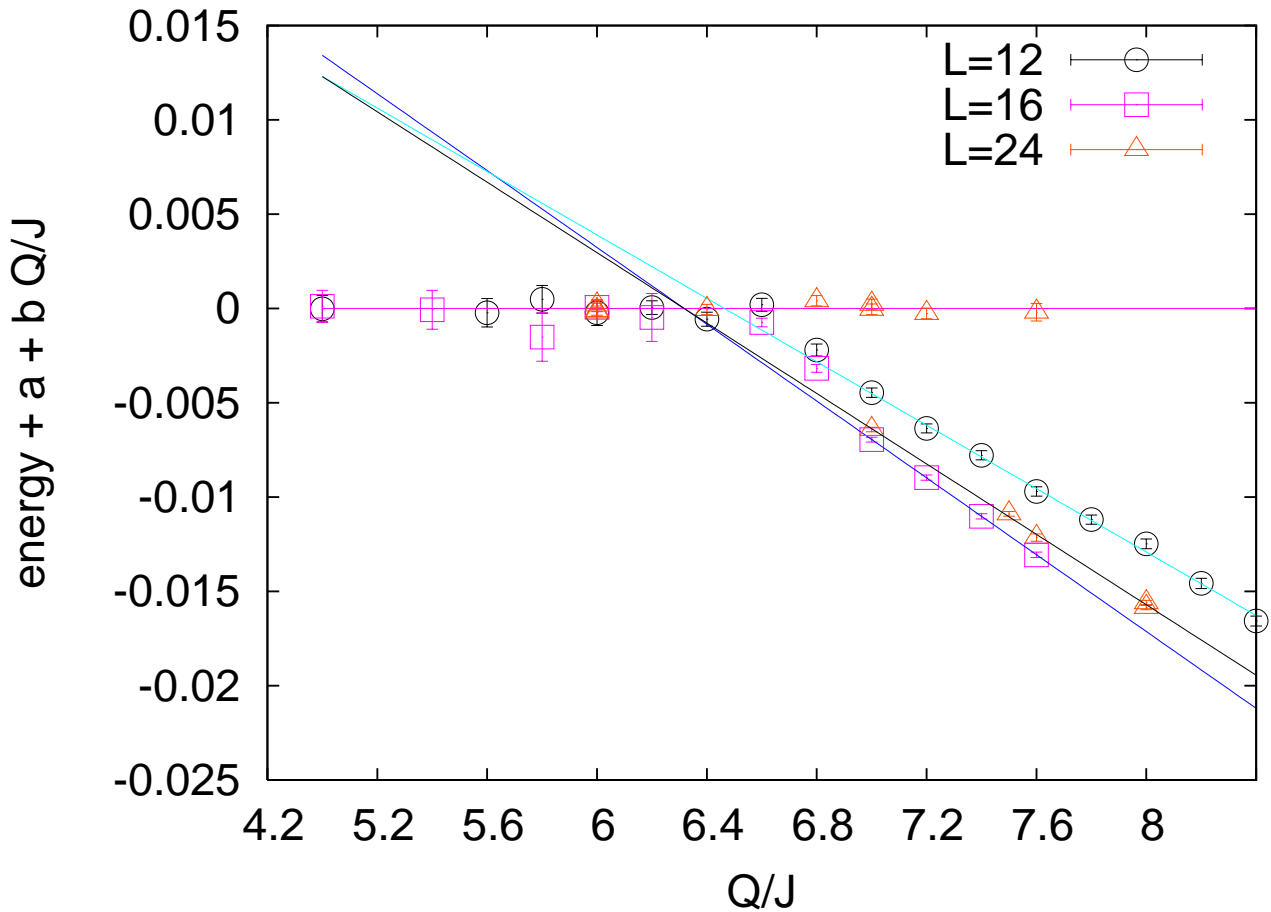}}
\caption{An avoided level crossing of the two competing candidate  ground states, 
 showing up as a kink in  the ground state energy of the honeycomb lattice $JQ$ model, 
when plotted as a function of the tuning parameter $Q$. This serves as clear signature 
of the first order transition taking place in this model system. Note that the
energy difference from the fitted linear behaviour on the antiferromagnetic side is plotted, 
 not the absolute ground state energies on both sides -  this  is just to highlight the discontinuity 
of slopes of the two curves at the transition point. To obtain linear fits to energy values on the valence
bond nematic side, 
the data points at $Q/J<6.8$ have not been considered.}
\label{el16}
\end{figure}

From our numerical study of these observables, we deduce that a  transition from the Neel phase to the valence bond 
nematic phase takes place as 
$Q/J$ is increased to the vicinity of $Q/J \sim 7 $. In this vicinity, 
there are clear first order jumps in both the order parameters as shown in 
Fig.\ref{orderparameterVsQ_typ0_stg}. At first sight, the data suggests that the transition point,
as determined from the location of the jump, may have a tendency to drift
with increasing system size.
However,  this is actually an artifact of the slowness of the code near the first order 
transition region at larger sizes, as is clear from an alternate
more precise analysis which we now summarize. In this alternative
approach, we determine the transition 
point by locating the kink in the plot of energy as function of tuning parameter
at smaller sizes, where there are no such issue of slowness of the Monte Carlo
update procedure used for calculating ground state properties. 
As shown in Fig.~\ref{el16}, the energy per site  shows a linear behaviour,
but with slopes of the line changing discontinuously near the transition. This can be understood as an avoided 
level crossing taking place at the first order transition point. Identifying the crossing point 
of these two linear extrapolations for the energy on is a convenient way to determine the transition point~\cite{Sen_Sandvik_2010} with reasonable accuracy. Using this method, 
with data from three different system sizes $L=12,16$ and $24$, we  identify the  first order transition to be at $Q/J = 6.4\pm 0.2$.

\begin{figure}[h]
{\includegraphics[width=\columnwidth]{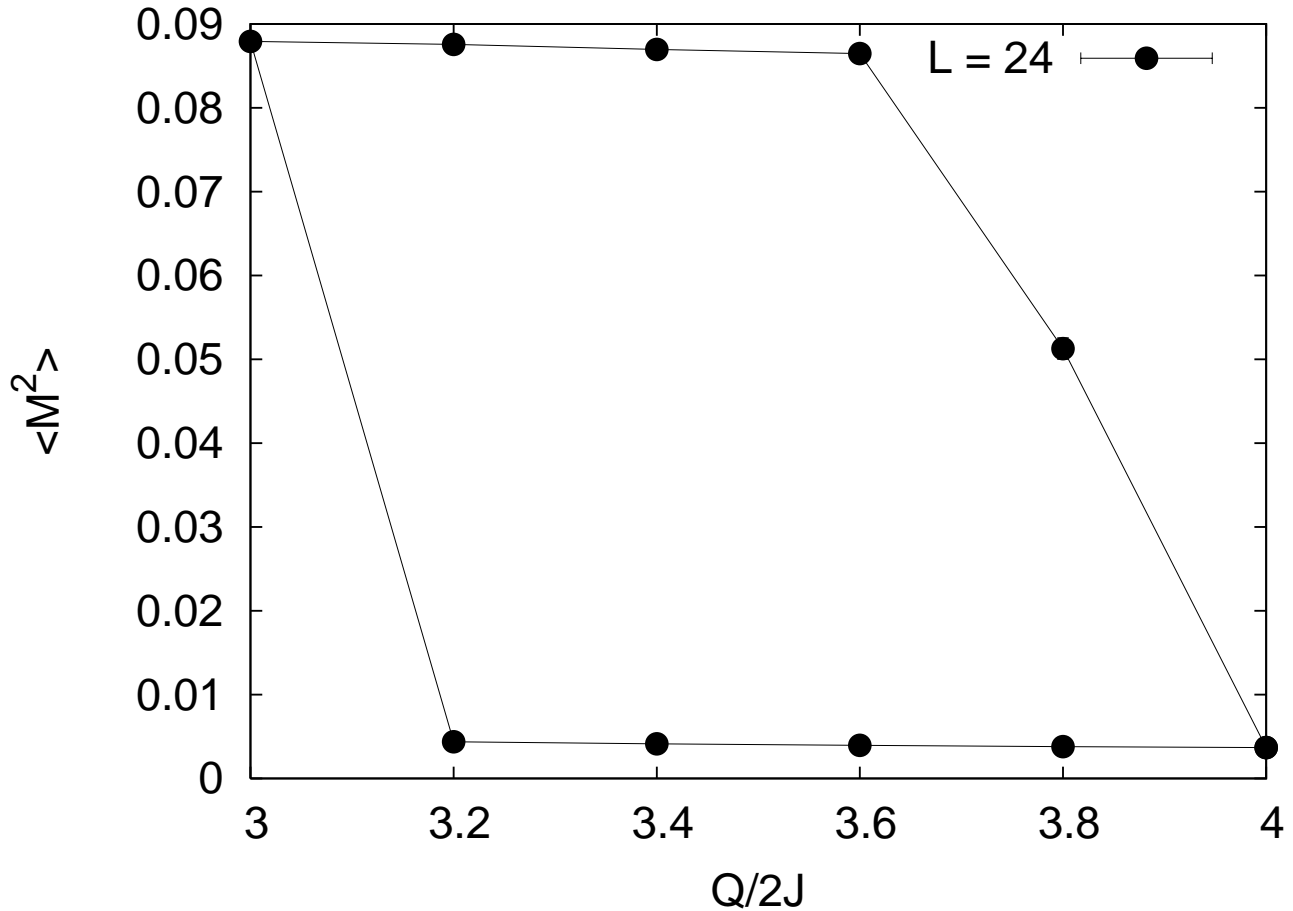}}
\caption{Hysteresis of $\langle {M}^2 \rangle$ around transition point for honeycomb lattice $JQ$ model. } 
\label{hysteresis_stg}
\end{figure}

First order transitions are associated with two coexisting  free energy minima 
corresponding to the two phases. This leads to the appearance of hysteresis effects in the vicinity of the transition for all but the most efficient
Monte Carlo algorithms when 
a configuration from one of the phases is evolved by tuning the coupling constant
across the critical value at finite rate (i.e without allowing for an extremely
large number of equilibriation steps between each successive change of coupling
constant). For example, starting with a configuration
deep in the valence bond nematic phase, the system remains stuck in
the metastable valence bond nematic state  even when $Q$ is reduced
to less than $Q_c$ and conversely, the system remains in the Neel
phase upon ramping $Q$ up beyond $Q_c$ if one starts with
a typical configuration obtained from the projection algorithm deep
in the Neel phase.
We present such a hysteresis plot for antiferromagnetic order parameter (Fig.\ref{hysteresis_stg}) 
near the transition to emphasize the first order nature of the transition. 

\begin{figure}[h]
{\includegraphics[width=\columnwidth]{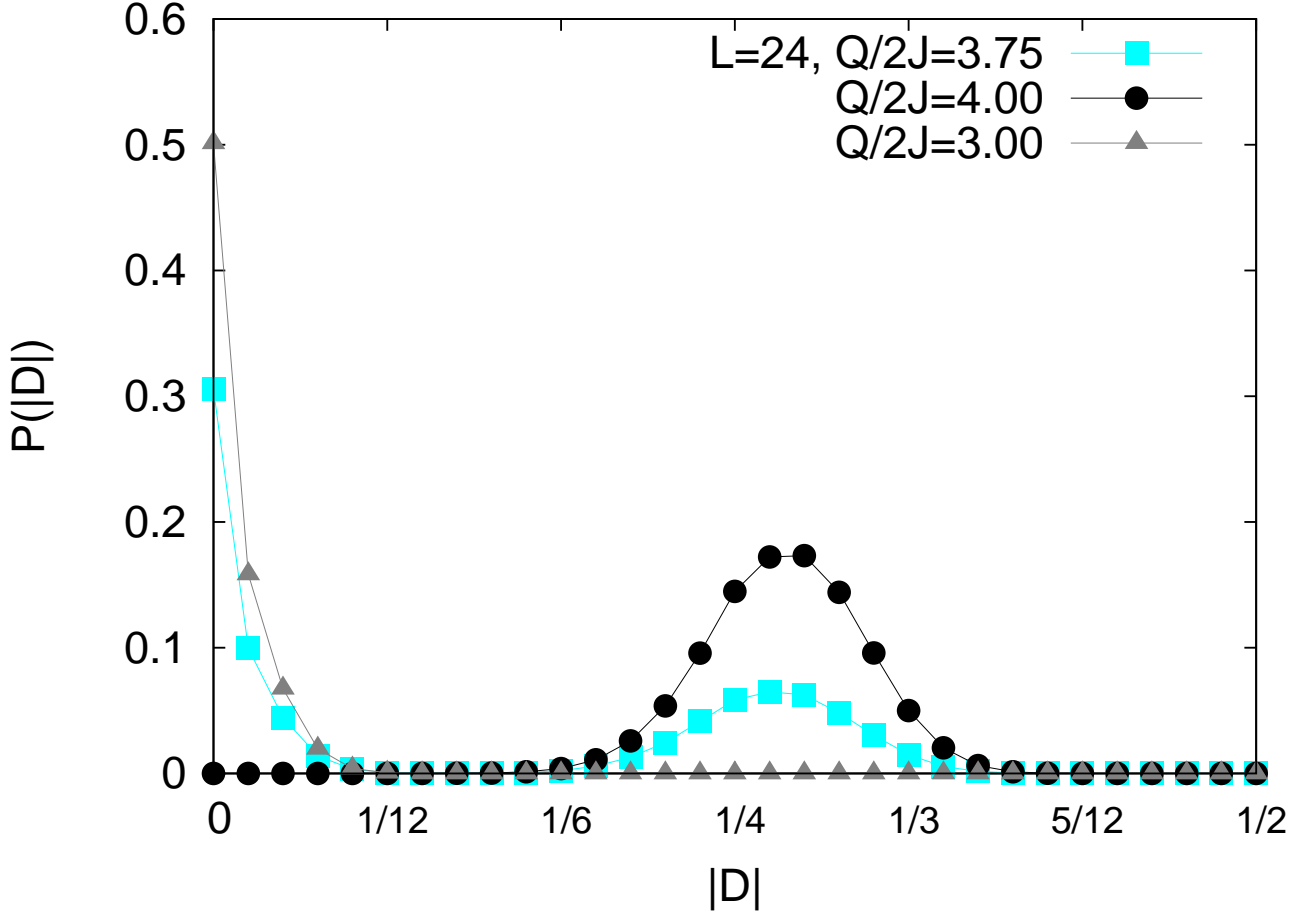}}
\caption{Histogram of valence-bond-nematic order parameter $|D| \equiv |D|(l)$ for L=24 at three
 different $Q$ values for honeycomb lattice JQ model. Presence of double peak structure
 at $Q/J = 7.5$ indicates first order nature of transition. } 
\label{histo_honeycomb}
\end{figure}

Our QMC simulations also yield a characteristic double peaked order parameter histogram near $Q_c$ as shown in
Fig.~\ref{histo_honeycomb}.
However, the `time' (number of Monte Carlo steps)  taken for the system to `tunnel' between the two phases is extremely large, and it
is therefore prohibitively expensive to perform runs that could
encompass a large number of such tunneling events. As a result,
it is not possible with our limited computational resources
to obtain accurate estimates of the relative heights of the two peaks in the histogram. 

\begin{figure}[h]
{\includegraphics[width=\columnwidth]{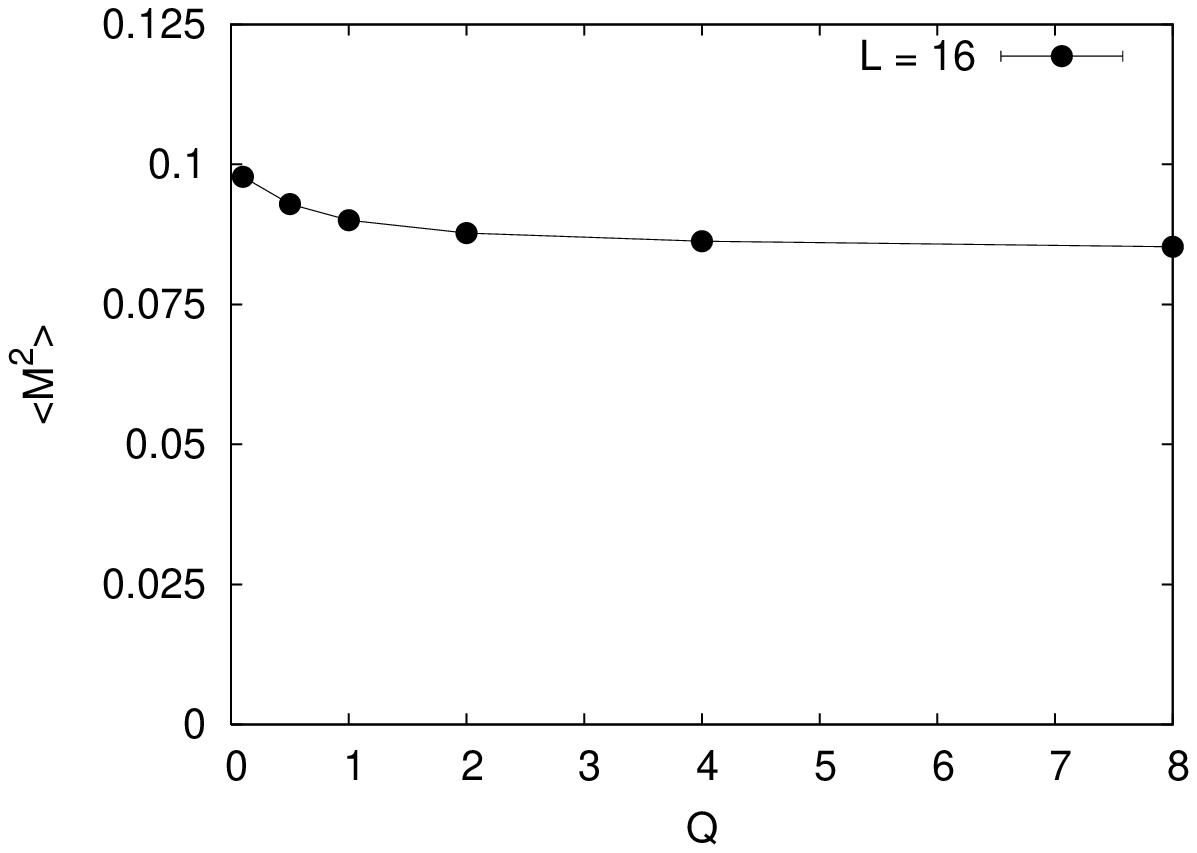}}
\caption{Order parameters $\langle M^2 \rangle$ as $Q/J$ is  varied for the $JQ$ honeycomb model
 with plaquettes made up of parallel bonds from neighbouring hexagons, showing no sign of 
a transition with increasing $Q/J$.} 
\label{orderparameterVsQ_typ1_stg}
\end{figure}
 Finally, we note that in a variant of staggered
$JQ_2$ model on honeycomb lattice, where parallel bonds from neighbouring hexagons make up 
the plaquette interactions, we find that increasing $Q$ fails to destroy AF order. For this model, as shown in
Fig.~\ref{orderparameterVsQ_typ1_stg}, there is no hint of a transition
even at large $Q/J$.

\section{Discussion}
\label{discussion}
We have thus studied an example of a transition between a Neel ordered antiferromagnet and a three-fold rotation
symmetry breaking valence bond nematic on the honeycomb lattice. Our numerical results
showed that the transition is strongly first order, consistent with the general scenario outlined
in the introduction. 
The transition point obtained from our analytical 
considerations is at $Q/\tilde{J} \approx 1.35$
($Q/J \approx 4.2$), which is in reasonable
agreement with the results of our QMC simulations,
which give a strong first order transition at $Q/\tilde{J} \approx 1.5$ ($Q/J \approx 6.4$). Furthermore,
the nearly $Q$ independent values of the staggered magnetization $M$ in
the N\'eel phase,
and of the nematic order $|D|$ in the staggered dimer state, are also consistent with our analytic
considerations. These observations suggest that the bond operator approach provides a 
reasonable account of the dimer phase in situations where the kinetic fluctuations of dimers
are not important in the vicinity of the transition. This appears to be different from the N\'eel
to columnar dimer transition in the square lattice case where such singlet fluctuations appear
to be important in driving the dimer order to zero at the quantum phase transition.\cite{Kotov}

\section{Acknowledgements}
We acknowledge computational resources at TIFR, grant support (KD) from the Indian DST (DST SR/S2/RJN-25/2006), NSERC of Canada (AP), and an Early
Researcher Award from the Government of Ontario (AP), and thank Arnab Sen and Anders Sandvik for useful discussions.

\end{document}